# Estimating Demand for Lamb, Beef, Pork, and Poultry in Canada

**Zakary Rodrigue Diakité[1,2]**

[1]Center for Research on the Economics of the Environment, Agri-Food, Transports, and Energy (CREATE), Laval University, Quebec, Canada

[2]Department of Agricultural Economics and Consumer Science, Laval University, Quebec, Canada

Email: zakary.diakite@icloud.com

## Abstract

This paper investigates the demand for lamb, beef, pork, and poultry in Canada, both at the national level and in disaggregated provinces, to identify meat consumption patterns in different provinces. Meat consumption plays a significant role in Canada's economy and is an important source of calories for the population. However, meat demand faces several consumption challenges due to logistic constraints, as a significant portion of the supply is imported from other countries. Therefore, there is a need for a better understanding of the causal relationships underlying lamb, beef, pork, and poultry consumption in Canada. Until recently, there have been no attempts to estimate meat consumption at the provincial level in Canada. Different Almost Ideal Demand System (AIDS) models have been applied for testing specifications to circumvent several econometric and theoretical problems. In particular, generalized AIDS and its Quadratic extension QUAIDS methods have been estimated across each province using the Iterative Linear Least Squares Estimator (ILLE) estimation Method. Weekly retail meat consumption price and quantity data from 2019 to 2022 have been used for Canada and for each province namely Quebec, Maritime provinces (New Brunswick, Nova Scotia, and Prince Edward Island), Ontario, total West (Yukon, Northwest Territory and Nunavut), Alberta, Manitoba-Saskatchewan and Manitoba as well as British Columbia. Consistent coefficients and demand elasticities estimates reveal patterns of substitution and/or complementarity between the four categories of meat. Meat consumption patterns differ across each province. Results show that the demand for the four categories of meat is responsive to price changes. Overall, lamb expenditure was found to be elastic and thus considered a luxury good during the study period, while the other three categories are considered normal goods across Canada.

## Keywords

Meat Demand; Estimation; Almost Ideal Demand System; Elasticity of Demand; Expenditure

## JEL classification Codes

C12 Hypothesis Testing: General;
C13 Estimation: General;
C51 Model Construction and Estimation;
D11 Consumer Economics: Theory;
D12 Consumer Economics: Empirical Analysis

# 1. Introduction

The meat industry is one of the key agricultural sectors in Canada, particularly in three provinces: Ontario, Quebec, and Alberta. These provinces lead the country in local livestock production. According to Statistics Canada, the Canadian government agency responsible for producing statistics, the local production of mutton and lamb meat for the three provinces combined accounted for 71% of the total Canadian production estimated to be 16,700 tonnes in January 2021. Of this, 131 tonnes were exported. Meanwhile, the total importation of mutton and lamb meat during the same period was 22,862 tonnes. In January 2021, Alberta alone produced 40% of the total local Canadian beef and veal meat estimated to be 1.384 million tonnes, of which 506,578 tonnes were exported and 161,513 tonnes were imported. The three leading provinces accounted for 81% of the total Canadian pork meat production, estimated to be 2.28 million tonnes, of which 1.44 million tonnes were exported and 252,124 tonnes were imported. In the same year, the Canadian commercial production of eviscerated chicken and turkey meat was estimated to be 1.45 billion kilograms. Poultry alone accounted for 1.3 billion kilograms, with 60.4% produced in Quebec and Ontario. Canadian local meat production is thus mainly concentrated in three provinces (Alberta, Quebec, Ontario). Despite local meat production, the livestock production sector in Canada remains dependent on international trade. Exportation and importation are combined with country-level logistic purposes that play a significant role in meat demand regulation in terms of quantity and quality.

Since the 1980s, global livestock products have progressively increased (Komarek et al., 2021) but a reduction in meat consumption in industrialised countries, including Canada marked the evolution of food consumption habits in recent years. As food consumption in general, meat consumption habits are driven by numerous factors that include income, age, financial status, price fluctuations, meat supply, socio-demographic characteristics, nutritional trends, taste preference and increasingly the awareness of the health hazards related to meat consumption (Moschini & Meilke, 1989; Capps Jr & Schmitz, 1991). In addition, environmental care led to reducing greenhouse gas (GHG) emissions through the consumption of vegetable protein instead of those meat. Vegetable-based diets have fewer negative impacts on the environment compared to animal-based diets (Komarek et al., 2021). These factors influence meat consumption and demand according to geographical locations.

Until recently, few studies have been conducted to analyse meat consumption patterns in various locations and understand the demand for meat. Additionally, no attempts have been made to estimate meat demand at a territorial level, especially province level for a specific country like Canada. Our objective is to conduct a meat demand estimation at the Canada and provincial levels to capture meat demand patterns, magnitude, and structural changes while taking location into account. Analyzing meat demand across Canada and its provinces is essential to understand Canadian meat demand patterns, while accounting for geographic and socio-demographic characteristics, and to understand the causal relationships underlying lamb, beef, pork, and poultry consumption in Canada.

To analyze meat consumption trends across Canadian provinces and measure consumption disparities between them, several extensions of the Almost Ideal Demand System (AIDS) specified by (Deaton & Muellbauer, 1980a, 1980b), the most popular of all demand systems, have been run for testing specification against Canadian weekly retails price and quantity data collected between August 24, 2019, to August 13, 2022 for four meats (Lamb, Beef, Pork and Poultry). Testing specification and Consistency analysis lead to the selection the most reliable extension of the generalized AIDS model to carry out estimations across each province using "Iterative Linear Least Squares Estimator" (ILLE) estimation Method provided by micEconAids packages (Henningsen, 2017) and available in R studio (Core Team R, 2018).

Meat prices and expenditure elasticities were estimated and analyzed to understand the four categories of meat demand in Canada at both the national and disaggregated province levels and to cross-check meat consumption patterns in the different provinces. In addition, of the status of luxurious product of lamb meat, our results confirm disparities in meat demand across provinces, which are linked to socioeconomic and demographic characteristics. The study thus adds a critical insight to the literature on meat consumption patterns in Canada. Through comprehensive analysis, it explores regional preferences, revealing the dynamic nature of dietary habits influenced by cultural, economic, and geographical factors.

This introduction has provided a general overview of meat production and consumption in Canada, as well as stating the research problem, objective, and a brief overview of the methodology and results. The Materials and Methods section will describe the demand estimation functional form, the estimation framework, the data, the methodology, and the estimation principle. A Results section will present the estimation outcomes with interpretations and analysis. A discussion will follow regarding demand estimation systems, results, and province-specific characteristics, before a conclusion in relation to our research objectives.

## 2. Methodology

### 2.1. Demand Estimation Functional Forms

The demand specifies the consumer's willingness and capacity to purchase a certain quantity of a specific good at the different possible prices according to his budget at a precise moment.

The demand is recognised to be influenced by factors such as the preferences of consumers between a panel of goods, prices of a substitute good, changes in conditions (Saison, pandemic, consumer, technology…), income level or variation. Several demand estimations functional forms based on consumer theories have been developed and used since the middle of the 20th century. The series of demand functional forms specification starts with the linear expenditure systems (LES) specified by Stone (Stone, 1954). The Rotterdam models (Barten, 1964; Theil, 1965; Theil, 1976) and then the translog model (Christensen et al., 1975) have followed and have been used until the specification of the Almost Ideal Demand System (AIDS) by (Deaton & Muellbauer, 1980a, 1980b) which has considerable advantages over the Rotterdam and translog models. AIDS was extended to its Quadratic form named the Quadratic Almost Ideal Demand System, QUAIDS (Banks et al., 1997; Poi, 2012). Before the extension of AIDS to QUAIDS, the Exact Affine Stone Index (EASI) was suggested in 2007 (Lewbel & Pendakur, 2008; Lewbel & Pendakur, 2009).

Since the specification of the Almost Ideal Demand System (AIDS) (Deaton & Muellbauer, 1980a, 1980b) and its Quadratic extension QUAIDS (Banks et al., 1997; Poi, 2012), they are the most popular demand estimation models used for empirical demand analysis. The Almost Ideal Demand System AIDS model (Deaton & Muellbauer, 1980a, 1980b) starts from a cost or expenditure function known as the PIGLOG ("Price Invariant Generalized Logarithmic") represented via the cost or expenditure function which defines the minimum expenditure necessary to attain a specific utility level at given prices. That function is denoted as $c(u,p)$ for utility $u$ and price vector $p$ is the general starting point for the specification of the AIDS model. That cost function is specified as follows (Deaton & Muellbauer, 1980a, 1980b).

$$\log c(u, p) = (1-u)\{a(p)\} + u \log\{b(p)\} \qquad (1)$$

Several desirable properties encourage the use of the Almost Ideal Demand System" (AIDS) often used in applied demand analysis. The most emphasized in the literature is that the AIDS model satisfies the *axioms of choice*, it *aggregates over consumers* and gives a *first-order approximation* to any demand system (Deaton & Muellbauer, 1980a, 1980b). Other properties completed these most emphasized properties to the complete list mentioned (Deaton & Muellbauer, 1980a, 1980b). AIDS models have a functional form consistent with known household-budget data, it is of great simplicity of estimation, it can be used to test the restrictions of *homogeneity* that suggest the absence of money illusion and *symmetricity* that assures consumer rationality through Slutsky`s symmetricity that is the fact that the compensated cross-price derivatives or elasticities are equal (Deaton & Muellbauer, 1980a, 1980b). The opposite to the Rotterdam and the translog model that have been extremely estimated before its specification, the AIDS model has the particularity to have all the desirable properties cited at the same time. In addition to homogeneity and symmetricity restriction, the plausibility of the demand estimation using AIDS, should also satisfy conditions from economic theory such as adding-up condition that justify the budget constraint and negativity condition that justify the concavity of expenditure function regarding price (Deaton & Muellbauer, 1980a, 1980b).

The Quadratic Almost Ideal Demand System QUAIDS (Banks et al., 1997; Poi, 2012) is a direct extension of the generalized almost ideal demand system AIDS. The quadratic extension of AIDS has the particularity to consider the existence of a non-linear Engel curve (Banks et al., 1997; Gostkowski, 2018). Together, the quadratic or generalized AIDS is recognised to be more accurate model (Blundell & Robin, 1999; Mizobuchi & Tanizaki, 2014).

In our research particular attention is given the generalised AIDS which will be the core of the nested demand systems investigated. The generalized Almost Ideal demand system is specified as follows.

$$w_{it} = \alpha_{it} + \sum_{j=1}^{n} \gamma_{ij} \ln p_{jt} + \beta_i \ln(X_t/P_t) \quad (2)$$

We define the Quadratic Almost Ideal Demand System QUAIDS here taking the formulation of (Denton & Mountain, 2004).

$$w_{it} = \alpha_{it} + \sum_{j=1}^{n} \gamma_{ij} \ln p_{jt} + \beta_i \ln(X_t/P_t) + \lambda_i \left(\ln(X_t/P_t)\right)^2 \quad (3)$$

where is $X_t$ is the total expenditure $w_{it}$ is the expenditure share associated with the $i^{th}$ good, $\alpha_i$ is the constant coefficient in the $i^{th}$ share equation, $\gamma_{ij}$ is the slope coefficient associated with the $j^{th}$ good in the $i^{th}$ share equation, $p_j$ is the price on the $j^{th}$ good. $\beta_i$ and $\lambda_i$ are coefficients associated with the $i^{th}$ share equation. The translog price index is specified as follows:

$$\ln P_t = \alpha_0 + \sum_{i=1}^{n} \alpha_i \ln p_{it} + \frac{1}{2} \sum_{i=1}^{n} \sum_{j=1}^{n} \gamma_{ij} \ln p_{ti} \ln p_{jt} \quad (4)$$

Which is the translog Price index where $\alpha_0$ corresponding to the "subsistence" expenditure (Asche & Wessells, 1997). The translog Price index is often replaced in their linear approximation of the AIDS model (LA/AIDS) by the stone price index below. The translog price index permits the systems to be nonlinear. When replaced by the Stones price index led to the LA/AIDS (linear approximation of the Demand system). Stone's price index is specified as follows:

$$\ln P^* = \sum_{i=1}^{n} w_i \ln P_i \quad (5)$$

Theoretically, the adding-up conditions are satisfied if $\sum_{i=1}^{n} \alpha_i = 1$, $\sum_{i=1}^{n} \beta_i = 0$, $\sum_{i=1}^{n} \gamma_{ij} = 0$

Homogeneity is satisfied exclusively for all $\sum_{j=1}^{n} \gamma_{ij} = 0$ and symmetry is satisfied if $\gamma_{ij} = \gamma_{ji}$.

## 2.2. Estimation Principle and Measure Unit

The almost Ideal Demand System (AIDS) can be estimated using R package

"micEconAids" proposed by (Henningsen, 2017). We use as (Henningsen, 2017) suggest the generalised nonlinear AIDS model, which is estimated using the Iterated Linear Least Square Estimation (ILLE) method. During the process, the share equations are estimated with linear techniques based on the ILLE estimations starting with the stone price index. The estimation starts with a fixed translog price index (corresponding to stone price index here) which is updated with the coefficients obtained from the previous step. Technically, by using the initials values of coefficients given by the linear approximation LA/AIDS estimation to calculate the translog price index, the expenditure share equation is estimated by linear estimation techniques holding the translog price index fixed and then the translog price index is updated with the newly estimated coefficients. This is repeated until the coefficients converge (Blundell & Robin, 1999; Henningsen, 2017).

One other simplification that was necessary to perform the study in accordance with its objective was to perform demand analysis and estimation at territory level instead of working at the household or individual level. Considering economic theory which suggests that the meat product requested from the grocery store depends on its own price, the prices of other meat products, and individual/family income (Tryfos & Tryphonopoulos, 1973), we aggregate quantities and revenues to define them at territory level.

### 2.3. Data Sources and Description

#### 2.3.1. Data Sources and Organisation

This research was based on store-level weekly data from seven Canadian administrative regions, notably Quebec (QC), Maritime (MA), Ontario (ON), Total West (WE), Alberta (AL), Manitoba-Saskatchewan (MS), and British Columbia (BC). The dataset used is from a private source provided by the market data collection specialist in Canada named Nielsen. Due to privacy and commercial considerations, it has not been directly disclosed in the study. Nevertheless, the data remains available upon request, and its relevant characteristics, related statistics, descriptions, and derived results are detailed in the current research. The data cover 95% of the entire Canada's weekly meat consumption purchased at the retail level from August 24, 2019, to August 13, 2022. The data are constituted with the four meat groups, including lamb, beef, pork as well as poultry. They include meat average prices (CA $/kg), meat expenditures (CA $), and meat quantities (kg). The expenditure value for each of the four considered meats is determined by multiplying its price by its corresponding quantity. For each product during the periods under consideration, the weekly expenditure shares are calculated as the ratios of the expenditure for each meat group to the total meat expenditure, which is determined by summing the weekly expenditures of the four meat groups. To perform a cross-comparison of meat demand at the province level, the total expenditure, the expenditure shares, and the quantities are determined at the province level. On the basis of the parameters described, an input matrix for estimation was constructed for each location.

### 2.3.2. Data Overview

The dispersion of meat consumption and prices around their average (coefficient of variation: CV) are in general more than 10% for all considered meat. Lamb and Poultry present the highest consumption (Quantities) and prices CV. In terms of consumption of lamb and poultry, CV are respectively 66.21% and 53.71% while those values are 17.01% and 29.40% for prices. Poultry prices are the most volatile followed by lamb prices. Overwise, Lamb consumption is the most volatile over time, followed by poultry. Beef and pork consumption and prices have the lowest CV close to 10% among the considered meats. (Table 1)

Meat presents seasonal pick of consumption over our study time frame. The pick of consumption is the same for the four meat groups considered. Canada's weekly meat consumption evolution for lamb, beef, pork and poultry from August 24, 2019, to August 13, 2022, shows that the more pronounced pick is located between the month of October and the end of January. We also have a small, pick between March and April. These periods include Christmas and religious celebrations. Canada's weekly meat consumption evolution from August 24, 2019, to August 13, 2022, for the four meats considered in this study confirms that lamb is the most expensive meat. Regarding Canada's weekly meat price ($/kg) data evolution from August 24, 2019, to August 13, 2022, Poultry and pork have slightly the same averages, with a higher volatility in the price of poultry, associated to a CV of 29.40%. Beef is the meat with intermediate prices between lamb and pork.

### 2.4. Demand Modelling and Estimations

### 2.4.1. Demand Nested Models

Four Almost Ideal Demand System (AIDS) nested models (Models that can be obtained from each other by imposing parametric restrictions or a limit of an approximation) were assessed against Canada provinces store-level weekly meat consumption data. We assume that plausible model that best fits the data at the country level (Canada) will be also plausible for Canada provinces and their data since provinces are respectively part of the country and their data as well. Four Nested AIDS models are constructed based on the scientific literature and our research objectives. The more complete (complicated) model call Model_1 is obtained by including 4 demand shifters that include the quadratic logged term and the seasonal effects to the generalised Almost Ideal Demand System specified in Equation (2). Regarding the fact that quadratic extension QUAIDS of the generalised AIDS model is not yet considered in the R package "micEconAids" proposed by (Henningsen, 2017), we integrated the quadratic term as a shifter variable in addition of a trigonometric terms that capture the seasonal effect in data. Model_1 is the most complete model assessed and have been integrated in our study to take account the maximum factors susceptible to influencing meat demand estimation. The theoretical representation of model_1 is specified in Equation (6).

Table 1. Meat (Lamb, beef, Pork, and Poultry) consumption and prices generals' statistic. SD = standard deviation; CV = coefficient of variation (%).

|  | Quantities (1000 kg) | | | | Price ($/kg) | | | |
|---|---|---|---|---|---|---|---|---|
|  | Lamb | Beef | Pork | Poultry | Lamb | Beef | Pork | Poultry |
| Minimum | 43 | 3.099 | 1.643 | 3.965 | 12.10 | 9.48 | 6.38 | 3.02 |
| Maximum | 398 | 7.448 | 4.620 | 21.474 | 26.53 | 19.51 | 11.19 | 11.14 |
| Mean | 81 | 4.364 | 2.510 | 6.656 | 20.13 | 14.45 | 9.12 | 7.35 |
| SD | 53.63 | 0.56 | 0.39 | 3.57 | 3.42 | 2.03 | 0.84 | 2.16 |
| CV (%) | 66.21 | 12.74 | 15.53 | 53.71 | 17.01 | 14.03 | 9.23 | 29.40 |

$$w_{it} = \alpha_i + \sum_{j=1}^{n} \gamma_{ij} \ln P_j + \beta_i \ln(X_t/P_t) + \lambda_i \left(\ln(X_t/P_t)\right)^2 \Big/ Q \\ + \alpha_i^c \cos\frac{2\pi t}{4} + \alpha_i^s \sin\frac{2\pi t}{4} + \alpha_i^t \cdot t \tag{6}$$

where $w_{it}$ is the expenditure share associated with the $i^{th}$ good, $\alpha_i$ is the constant coefficient in the $i^{th}$ share equation, $\gamma_{ij}$ is the slope coefficient associated with the $j^{th}$ good in the $i^{th}$ share equation, $p_j$ is the price on the $j^{th}$ good. $\beta_i$ and $\lambda_i$ are coefficient associated to the $i^{th}$ share equation, $Q$ the Cobb-Douglas aggregate price defined as follow $Q = \prod_{j=1}^{n} p_j^{\beta_i}$.

As $\beta_i$ appear simultaneously in (6) and the Cobb Douglas aggregate price expression, we use the coefficient $\beta_i$ estimated in Equation (2) to calculate $Q$ in the quadratic term of Model_1. To incorporate seasonality and trend, we augmented the model with the following trigonometric and a time trend expression $\alpha_i^c \cos\frac{2\pi t}{4} + \alpha_i^s \sin\frac{2\pi t}{4} + \alpha_i^t \cdot t$.

The second model assessed call Model_2 is a reduction of Model_1 by taking off from the equation the following quadratic term $\left(\ln(X_t/P_t)\right)^2 \Big/ Q$. It corresponds to generalized Almost Ideal Demand System (2) augmented by incorporating seasonality and trend through the trigonometric variables and a time trend variable. The theoretical representation of Model_2 is the following (7).

$$w_{it} = \alpha_i + \sum_{j=1}^{n} \gamma_{ij} \ln P_j + \beta_i \ln(X_t/P_t) + \alpha_i^c \cos\frac{2\pi t}{4} + \alpha_i^s \sin\frac{2\pi t}{4} + \alpha_i^t \cdot t \tag{7}$$

The third model called Model_3 is an extension of the generalized Almost Ideal Demand System specified in Equation (2) with integration of the following logged total expenditure term $\log(X_t)$ as an instrumental variable. The theoretical formulas of the Model_3 is specified in Equation (8)

$$w_{it} = \alpha_i + \sum_{j=1}^{n} \gamma_{ij} \ln P_j + \beta_i \ln(X_t/P_t) + \log(X_t) \tag{8}$$

The fourth model call Model_4 is simply the generalized Almost Ideal De-mand System specified in Equation (2) as follow

$$w_{it} = \alpha_{it} + \sum_{j=1}^{n} \gamma_{ij} \ln p_{jt} + \beta_i \ln(X_t/P_t) \quad (9)$$

The different nested models fit the generalized Almost Ideal Demand System AIDS proposed by (Deaton & Muellbauer, 1980a, 1980b). (Table 2)

### 2.4.2. Models' Assessment

Using in R studio (Core Team R, 2018) the "micEconAids" R package proposed by (Henningsen, 2017), we estimated the four nested models constructed and after analysing the R-squared Values of expenditure shares and quantities estimated, the satisfaction of the Monotonicity and the Concavity of the expenditure function regarding prices were checked to assess the plausibility of the models with the demand theory. We then performed models Likelihood Ratio Tests that is used to determine if two models are significantly different to choose the one that significatively best fit the data. We perform a likelihood ratio test which uses the following null and alternative hypotheses: the null hypothesis stipulate that the full model and the nested model fit the data equally well. In this case, we should use the nested (simple) model. The alternative hypothesis stipulate that the full model fits the data significantly better than the nested model. In this case, we should use the full (most complicated) model. As the threshold P value is usually set at 0.05 in our study; if the P value is high than 0.05, the full model and the nested model are not significatively different and fit the data equally well. Thus, we should use the nested model because the additional term of variables in the full model does not offer a significant improvement in accordance with the data. If the P value is smaller than 0.05, the null hypothesis is wrong, the full model and the nested model are significatively different, and the additional term of variables in the full model thus offers a significant improvement in accordance with the data. In this case, we accepted the more complicated model. Based on this analysis the most plausible model is identified. For the demand estimation and elasticities determination we used the most plausible model identified after the nested model's assessment. The model was run to estimate the four meat (lamb, beef, pork, and poultries) elasticities (expenditure, own price, and cross-price elasticities).

Table 2. Nested models functional forms.

| | Functional forms |
|---|---|
| Model_1 | $w_{it} = \alpha_i + \sum_{j=1}^{n} \gamma_{ij} \ln P_j + \beta_i \ln(X_t/P_t) + \lambda_i (\ln(X_t/P_t))^2 / Q$ $+ \alpha_i^c \cos \frac{2\pi t}{4} + \alpha_i^s \sin \frac{2\pi t}{4} + \alpha_i^t \cdot t$ |
| Model_2 | $w_{it} = \alpha_i + \sum_{j=1}^{n} \gamma_{ij} \ln P_j + \beta_i \ln(X_t/P_t) + \alpha_i^c \cos \frac{2\pi t}{4} + \alpha_i^s \sin \frac{2\pi t}{4} + \alpha_i^t \cdot t$ |
| Model_3 | $w_{it} = \alpha_i + \sum_{j=1}^{n} \gamma_{ij} \ln P_j + \beta_i \ln(X_t/P_t) + \log(X_t)$ |
| Model_4 | $w_{it} = \alpha_{it} + \sum_{j=1}^{n} \gamma_{ij} \ln p_{jt} + \beta_i \ln(X_t/P_t)$ |

### 2.4.3. Elasticity of Demand

Elasticities are the most important results on demand analysis, the **expenditure** and Marshallian (uncompensated) own and cross-price elasticities of the four-meat demand will be the main result of our research. In general, elasticities inform about the percentage change in demand in response to a one percent (marginal) change in prices or consumer income. Expenditure elasticities indicate how a marginal increase of the income affects the quantities consumed, these elasticities reveal how much the quantity consumed changes because of a marginal change in the total expenditure of meat products.

Expenditure elasticities will permit us to identify which one is necessity, luxury, inferior or even the normal meat among the four considered. When the expenditure elasticity is larger than one, the good is luxury and changes in quantity demanded are larger than those in expenditures. In the case that the elasticity is between zero and one, the meat will be classified as a necessity. Nevertheless, meat classified as luxuries and necessities remains part of normal meat. Inferior meat will be characterised by negative expenditure elasticity.

Cross-price elasticity reveals complementarity between two meats in the case when the elasticity is negative and substitution between two meats when their cross-price elasticity is positive. Meat own-price elasticity permit to appreciate the change in demand when the considered meat own price changes. Due to negativity restrictions and absence of Giffen good for which rises in price increase the demand, own-price elasticities are negative along our research. These elasticities have been cross checked and analysed for each Canadian province. Although our research elasticities are determined using R package "micEconAids" proposed by (Henningsen, 2017), the theoretical formulas used to estimate the elasticities are stated in equations (9) and (10). The Marshallian (uncompensated) own and cross price elasticity coefficients from generalised AIDS corresponding to model (2) are given by the following formula (9) (Asche & Wessells, 1997; Buse & Chan, 2000).

$$\varepsilon_{ijt} = -\delta_{ij} + \frac{\gamma_{ij}}{w_{it}} - \frac{\beta_i}{w_{it}} \left( \alpha_j + \sum_{j=1}^{n} \gamma_{ij} \ln p_{jt} \right) \tag{9}$$

where $\delta_{ij}$ is the Kronecker's delta; $\delta_{ij} = 1$ if $i = j$; $\delta_{ij} = 0$ if $i \neq j$

The expenditure elasticity for the generalised AIDS model is given by the following formula (10) (Asche & Wessells, 1997; Buse & Chan, 2000).

$$\eta_{ij} = 1 + \frac{\beta_i}{w_{it}} \tag{10}$$

## 3. Results

### 3.1. Demand Nested Models' Assessment

After estimating the demand using the four nested models against Canada data, the most plausible and convenient model with data was identified by analysing R-Squared values (of expenditures shares and quantities) and the likelihood ratio tests results. While adding up (budget constraint), homogeneity (absence of money illusion), and symmetry (slutsky symmetry) conditions were imposed to the four models, all of them have equally fulfilled monotonicity and concavity condition at 100% for Canada data. By fulfilling the concavity of expenditure regarding to prices which condition determine the plausibility of a demand system, the four Models satisfy the negativity condition when we used Canada data for estimations. (Table 3)

Except for Model_1, the three nested models (Model_2, Model_3 and Model_4) expenditure and quantities R-square values are respectively closer to each other's (Table 3). R squared values stand between 44% and 45% for lamb expenditure shares and are equal to 72% for lamb quantities. In beef case, R-Squared values are between 25% and 29% for expenditure and between 47% and 49% for quantities. In pork case R square values are between 30% and 33% for expenditure and 31% and 36% for quantities. For the four models' poultry quantities R-squared value is quite similar and corresponds to the highest R squared (between 88% and 89%). Model_1 (lamb, beef, and pork) R squared values for expenditure and quantities are not included in the previous intervals in addition to be the lowest in each case, we have a negative value in the case of quantities R squared value for pork. In the case on Model_1, for pork the quantity R-squared value is -157 while those of the three other models are between 30% and 35%. Model_1 globally present the worse R-square values compared to those of the three other models. Regarding R-squared value Model_1 seem to be to less accurate model in terms of expenditure and quantities estimations. Considering that Model_1 is less accurate, to identify among the four models, the most convenient and accurate model with the data, the likelihood ratio tests results was performed. (Table 4)

Regarding the Likelihood Ratio Test result in Table 4, the null hypothesis is

Table 3. Nested Models expenditure shares quantities R-squared Values. Es = Expenditure shares (%), Qt = Quantity (%)

| Es (%) or Qt (%) | Model_1 | | Model_2 | | Model_3 | | Model_4 | |
|---|---|---|---|---|---|---|---|---|
| | Es | Qt | Es | Qt | Es | Qt | Es | Qt |
| lamb | 45 | 56 | 45 | 72 | 44 | 72 | 44 | 72 |
| beef | 29 | 16 | 29 | 47 | 25 | 49 | 26 | 49 |
| pork | 34 | -157 | 33 | 31 | 31 | 36 | 30 | 35 |
| poultry | 29 | 89 | 22 | 88 | 28 | 88 | 28 | 88 |

**Table 4.** The likelihood ratio test result of the four nested models (Model_1, Model_2, Model_3, Model_4).

|         | #Df | LogLik | Df | Chisq   | Pr (>Chisq) |
|---------|-----|--------|-----|---------|-------------|
| Model 1 | 30  | 1379.4 |     |         |             |
| Model 2 | 27  | 1376.6 | −3  | 5.5970  | 0.1330      |
| Model 3 | 18  | 1369.9 | −9  | 13.4917 | 0.1416      |
| Model 4 | 18  | 1369.8 | 0   | 0.1496  | <2e−16***   |

"***" Significant at less than the 0.1% confidence level, "**" Significant at the 0.1% confidence level, "*" Significant at the 1% confidence level, "." Significant at the 5% confidence level, (blank) Significant at the 10% confidence level.

**Table 5.** The likelihood ratio test result for the nested models (Model_2 and Model_3).

|         | #Df | LogLik | Df | Chisq  | Pr (>Chisq) |
|---------|-----|--------|-----|--------|-------------|
| Model 2 | 27  | 1376.6 |     |        |             |
| Model 3 | 18  | 1369.9 | -9  | 13.492 | 0.1416      |

"***" Significant at less than the 0.1% confidence level, "**" Significant at the 0.1% confidence level, "*" Significant at the 1% confidence level, "." Significant at the 5% confidence level, (blank) Significant at the 10% confidence level.

wrong for the generalized Almost Ideal demand system (Model_4) against the more completed model (Model_1). Among these 2 models we accept the more complete model (Model_1) that fits the data significatively better than the nested model_4. Other way the null hypothesis is true for the model_2 and model_3, these two models fit the data equally well as the completed model Model_1. The nested models (model_2 and model_3) are thus preferred to the completed model.

To choose between the two nested models that fit better the data, we perform a second likelihood ratio test based exclusively on them (Model_2 and Model_3).

Given the second Likelihood Ratio Test result in Table 5, the null hypothesis is validated. The more complicated model in this case (Model_2) and the nested model here (Model_3) fit the data equally well. We should use the nested model (Model_3) because the additional term of variables in the full model do not offer a significant improvement in fit. The R-Square value analysis and the likelihood ratio test result led to use the model_3 as the most convenient with the data to determine the elasticity coefficient along the study (Table 4 & Table 5).

### 3.2. Provinces Level Demand Estimation Consistency and Accuracy

The most convenient Model_3 provides the expenditure shares and quantities R-squared detailed in Table 6. While adding up (budget constraint), homogeneity (absence of money illusion), and symmetry (slutsky symmetry) conditions were imposed, the most convenient Model_3 fulfilled monotonicity and concavity condition at 100% for Canada and its provinces as well. Negativity condition is thus fulfilled, and the demand system plausibility is proved and confirmed for the considered provinces.

Except for Maritime (Ma) and Manitoba-Saskatchewan and Manitoba (MS) lamb R-Squared values (respectively −233 and −83 for Ma and MS) that are negative, all the provinces expenditure shares R-squared values for the four considered meats are positive (Table 6 and Figure 1 & Figure 2). In the case of quantities, lamb R-squared values for Ma and MS are the lowest compared to those of other provinces.

R-Squared values are higher with more dispersed values for Quantities (Qt) at the opposite of expenditure shares (Es) cases. Quantities R-Squared values remain positive except for Maritime (Ma) lamb quantities R-Squared value (−122).

Table 6. R-squared values of expenditure shares (Es) and Quantities (Qt) for Canada (Ca), Quebec (Qc), Maritime (Ma), Ontario (On), Total West (We), Alberta (Al), Manitoba-Saskatchewan and Manitoba (MS) and British Columbia (BC).

|        | Items   | Ca | Qc | Ma   | On | We | Al | MS  | BC |
|--------|---------|----|----|------|----|----|----|-----|----|
| Es (%) | lamb    | 44 | 49 | −233 | 39 | 2  | 44 | −83 | 20 |
|        | beef    | 25 | 16 | 30   | 52 | 33 | 17 | 32  | 57 |
|        | pork    | 31 | 8  | 17   | 25 | 34 | 27 | 45  | 35 |
|        | poultry | 28 | 4  | 42   | 47 | 33 | 27 | 35  | 49 |
| Qt (%) | lamb    | 72 | 61 | −122 | 67 | 54 | 73 | 32  | 58 |
|        | beef    | 49 | 79 | 14   | 48 | 51 | 67 | 67  | 44 |
|        | pork    | 35 | 50 | 60   | 37 | 44 | 47 | 80  | 40 |
|        | poultry | 88 | 29 | 91   | 94 | 90 | 88 | 48  | 90 |

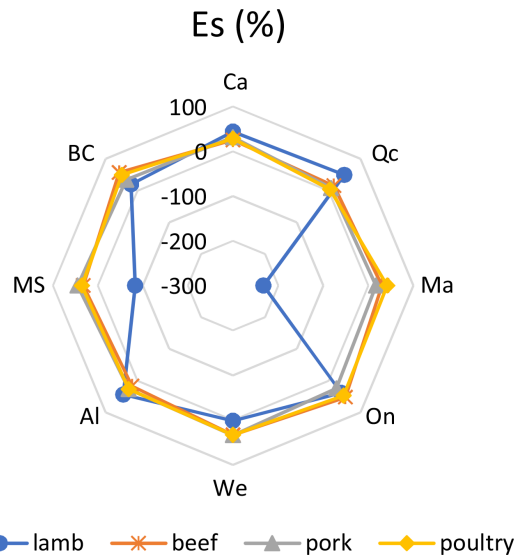

Figure 1. Expenditure shares (Es) R-Squared Values. (Canada (Ca), Quebec (Qc), Maritime (Ma), Ontario (On), total West (We), Alberta (Al), Manitoba-Saskatchewan and Manitoba (MS) and British Columbia (BC)).

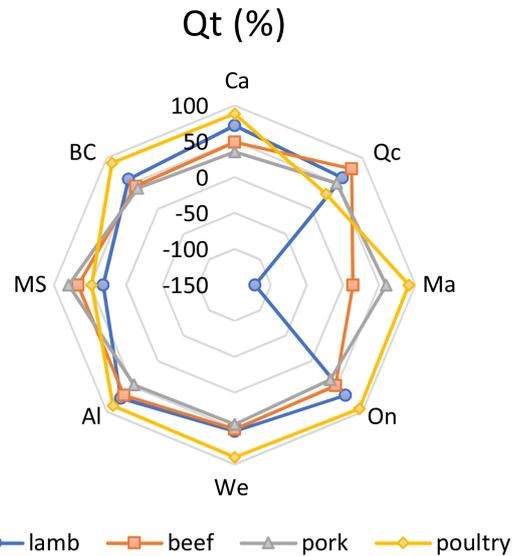

**Figure 2.** Quantities (Qt) R-Squared Values. (Canada (Ca), Quebec (Qc), Maritime (Ma), Ontario (On), total West (We), Alberta (Al), Manitoba-Saskatchewan and Manitoba (MS) and British Columbia (BC)).

Comparing Provinces R-squared values to each other, we cannot deduce that they have a pronounced difference in terms of quantities and Expenditures shares estimation accuracy except for Ma and MS provinces that have the lowest R-squared value specialty in lamb case.

### 3.3. Expenditure and Price Elasticities Cross Analysis

The results are organised around three categories of parameters estimated, including the Marshallian (uncompensated) own-price elasticities the expenditure elasticities and the Marshallian cross-price elasticities. According to the theory of economics such as the concavity of expenditure function regarding prices, the negativity of the price effect is reflected by a negative (uncompensated) Marshallian own-price elasticity for all meat. All meat's own price elasticities are also negative (Table 7), that is the case for all provinces considered in the study. Taking account, the means of elasticities value of provinces, Lamb and poultry own price elasticities permit to characterise these goods as elastic while beef and pork are not Table 9. As lamb and poultry are elastics those meat consumption varies quickly than their prices at provinces and country level Table 7.

The expenditure elasticities that inform about the status of a product is the second important parameter considered in that demand analysis. Canada country level Lamb and Poultry Expenditure Elasticities are respectively of 1.019 and 1.499 and the investigated provinces expenditure elasticities mean values are respectively 1.606 and 1.389 for Lamb and Poultry.

Lamb and Poultry expenditure elasticities are mostly significant and greater than one at country level, both meats are luxurious product what is not the case for Lamb meat in Alberta (Al), where Lamb meat is a necessity with a value of

Table 7. Marshallian (uncompensated) price and expenditure elasticities. (Canada (Ca), Quebec (Qc), Maritime (Ma), Ontario (On), total West (We), Alberta (Al), Manitoba-Saskatchewan and Manitoba (MS) and British Columbia (BC)).

| Quantities | Locations | Marshallian (uncompensated) Price Elasticities | | | | Expenditure Elasticities |
| --- | --- | --- | --- | --- | --- | --- |
| | | lamb | beef | pork | poultry | |
| lamb | Ca | −2.639*** | 0.138 | 1.812 | −0.330** | 1.019 |
| | Qc | −6.063** | 4.448 | −0.330 | 0.379 | 1.567 |
| | Ma | −2.890 | −0.390 | 1.793 | 0.107 | 1.380 |
| | On | −2.581*** | −0.329 | 1.440 | −0.132 | 1.603 |
| | We | −2.900** | −0.693 | 2.434 | −0.011 | 1.170 |
| | Al | −3.099 | 2.174 | 0.251 | −0.144 | 0.819 |
| | MS | −2.407 | −2.714 | −0.526 | 2.174 | 3.473 |
| | BC | −2.037* | −0.649 | 1.607 | −0.153 | 1.233 |
| beef | Ca | 0.005 | −0.650*** | −0.120** | −0.043. | 0.808*** |
| | Qc | 0.044 | −1.383*** | 0.145** | 0.113. | 1.081*** |
| | Ma | −0.0008 | −0.827*** | 0.043 | 0.066 | 0.719*** |
| | On | 0.004 | −0.612*** | −0.047 | 0.054. | 0.601*** |
| | We | −0.010 | −0.604*** | −0.108*** | 0.006 | 0.716*** |
| | Al | 0.030 | −1.043*** | 0.098* | 0.077** | 0.838*** |
| | MS | −0.015 | −1.315*** | 0.123*** | 0.442*** | 0.765*** |
| | BC | −0.009 | −0.537*** | −0.075. | 0.122*** | 0.494*** |
| pork | Ca | 0.123 | −0.214* | −0.472** | 0.008 | 0.555*** |
| | Qc | −0.002 | 0.380*** | −1.198*** | −0.071 | 0.891*** |
| | Ma | 0.072 | 0.196. | −0.722*** | −0.115. | 0.570*** |
| | On | 0.168* | −0.113 | −0.779*** | 0.120* | 0.605*** |
| | We | 0.143. | −0.277** | −0.666*** | 0.119* | 0.681** |
| | Al | 0.014 | 0.475*** | −1.108*** | 0.074. | 0.545*** |
| | MS | −0.006 | −0.273* | −1.697*** | −0.029 | 2.006*** |
| | BC | 0.146 | −0.345** | −0.685* | 0.127 | 0.756. |
| poultry | Ca | −0.016 | −0.392*** | −0.158*** | −0.932*** | 1.499*** |
| | Qc | 0.015 | 0.474* | −0.127 | −1.264*** | 0.902*** |
| | Ma | 0.0002 | −0.382*** | −0.267*** | −1.042*** | 1.691*** |
| | On | −0.006 | −0.345*** | −0.116. | −1.105*** | 1.571*** |
| | We | −0.004 | −0.385*** | −0.090 | −1.068*** | 1.547*** |
| | Al | −0.008 | −0.219* | −0.114*** | −1.163*** | 1.504*** |
| | MS | 0.043 | 0.618*** | 0.222*** | −1.618*** | 0.735** |
| | BC | −0.014 | −0.441*** | −0.106 | −1.215*** | 1.775*** |

"***" Significant at less than the 0.1% confidence level, "**" Significant at the 0.1% confidence level, "*" Significant at the 1% confidence level, "." Significant at the 5% confidence level, (blank) Significant at the 10% confidence level.

expenditure elasticity lower than one (Table 7).

Poultry is also a luxurious meat across Canada, provinces except in Quebec (QC) and Manitoba-Saskatchewan and Manitoba (MS) where the poultry meat is a necessity product given is expenditure elasticities less than one (Table 7). We have the highest expenditure elasticities of the research in the province of MS for Lamb and pork. Lamb meat has the status of elastic and luxurious good across Canada with an expenditure elasticity great than one for all provinces except in Alberta (Al) where the Expenditure elasticity reached 0,819. Lamb meat has a status of necessity product in the province of Alberta (Al).

In the area of MS Pork is a luxurious meat with the second highest value of expenditure elasticity (after Lamb expenditure elasticity for MS) (Table 7) at the opposite of other provinces in Canada where pork is a necessity product. Globally all considered meats in that study have been classified has luxurious or necessity thus normal meat. None of the considered meats have been identified as an inferior good given the positivity of their expenditure elasticity values.

Beef meat is characterised as a luxurious meat in Quebec (QC) with the highest value of expenditure elasticity, great that one at the opposites of other provinces where beef is a necessity. Beef and Pork are a booth necessity in Canada in general that is not the case in the province of MS where pork expenditure elasticity is 2.006 which characterises pork as a luxury product in MS. Thus, a one percent increase in expenditure leads to a rise in the consumption of pork meat by about 2.006 percent on average (ceteris paribus).

In the study, the maximum expenditure elasticity for lamb is associated with the province of MS and the minimum with Alberta (Al). The maximum beef expenditure elasticity is associated with the province of Quebec (QC) and the minimum to British Columbia (BC). The maximum pork expenditure elasticity is associated with the province of MS and the minimum to Alberta (Al) and the maximum poultry expenditure elasticity is associated with BC and the minimum to MS. In the provinces of MS, BC, and Al meat demand is more sensible to expenditure variation.

The other important parameter analysed is the Marshallian cross-price elasticities. Based on the Marshallian cross-price elasticities, in the provinces of QC and MS and at the opposite of the other provinces the couple Lamb and pork have the status of complementary goods. Beef and poultry are substitute good in QC and MS at the opposite of the other provinces where they are complementary goods.

When the Lamb price increase it is substituted with pork and then beef. When beef and pork price increases, they are substituted by lamb meat. When the poultry price increase it is substituted with pork meat (Table 8). These substitutions are in general function of meat prices evolution in addition of the location socio-economic and demographic characteristic.

Regarding the cross-price elasticities, In Canada, substitution status is associ-ated with the following couple of goods Lamb-Beef, Lamb-Pork, Beef-Lamb, Pork-Lamb, and Poultry-Pork. Among these couples of goods only Pork-Lamb is elastic given its cross-price elasticity value of 1.812. Furthermore, the following couples of meats, Lamb-Poultry, Beef-pork, Beef-poultry, Pork-Beef, Pork-Poultry, Poultry-Lamb, Poultry-Beef have the status of complementary goods given the negative sign of their cross-elasticity values. Lamb meat is elastic across Canada and its often substituted with beef or pork because their cross-price elasticities are positive, but the substitution magnitude is low in both cases. Poultry has a negative cross-price elasticity with lamb, it is therefore used as a complement of lamb meat in Canada country level. (Table 9)

Table 8. Marshallian (uncompensated) price and expenditure elasticities for Canada. Canada (CA), Minimum (Min), average (Mean), Maximum (Max), Standard Deviation (SD).

| Quantities | Marshallian (uncompensated) Price Elasticities | | | | Expenditure Elasticities |
| --- | --- | --- | --- | --- | --- |
| | lamb | Beef | Pork | Poultry | |
| Lamb | −2.639*** | 0.138 | 1.812 | −0.330** | 1.019 |
| Beef | 0.005 | −0.650*** | −0.120** | −0.043. | 0.808*** |
| Pork | 0.123 | −0.214* | −0.472** | 0.008 | 0.555*** |
| Poultry | −0.016 | −0.392*** | −0.158*** | −0.932*** | 1.499*** |

"***" Significant at less than the 0.1% confidence level, "**" Significant at the 0.1% confidence level, "*" Significant at the 1% confidence level, "." Significant at the 5% confidence level, (blank) Significant at the 10% confidence level.

Table 9. Marshallian (uncompensated) price and expenditure elasticities summary. Canada (CA), Minimum (Min), average (Mean), Maximum (Max), Standard Deviation (SD).

| | Summary | Lamb | Beef | Pork | Poultry | Expenditure |
| --- | --- | --- | --- | --- | --- | --- |
| Lamb | Min | −6.063 | −2.714 | −0.526 | −0.153 | 0.819 |
| | Mean | −3.140 | 0.264 | 0.953 | 0.317 | 1.606 |
| | Ca | −2.639 | 0.138 | 1.812 | −0.330 | 1.019 |
| | Max | −2.037 | 4.448 | 2.434 | 2.174 | 3.473 |
| | SD | 1.337 | 2.329 | 1.147 | 0.841 | 0.865 |
| Beef | Min | −0.015 | −1.383 | −0.108 | 0.006 | 0.494 |
| | Mean | 0.006 | −0.903 | 0.026 | 0.126 | 0.745 |
| | Ca | 0.005 | −0.650 | −0.120 | −0.043 | 0.808 |
| | Max | 0.044 | −0.537 | 0.145 | 0.442 | 1.081 |
| | SD | 0.022 | 0.350 | 0.102 | 0.145 | 0.186 |
| Pork | Min | −0.006 | −0.345 | −1.697 | −0.115 | 0.545 |
| | Mean | 0.076 | 0.006 | −0.979 | 0.032 | 0.865 |
| | Ca | 0.123 | −0.214 | −0.472 | 0.008 | 0.555 |

Continued

|  |  |  |  |  |  |  |
|---|---|---|---|---|---|---|
|  | Max | 0.168 | 0.475 | −0.666 | 0.127 | 2.006 |
|  | SD | 0.076 | 0.339 | 0.381 | 0.102 | 0.517 |
|  | Min | −0.014 | −0.441 | −0.267 | −1.618 | 0.735 |
|  | Mean | 0.004 | −0.097 | −0.085 | −1.211 | 1.389 |
| Poultry | Ca | −0.016 | −0.392 | −0.158 | −0.932 | 1.499 |
|  | Max | 0.043 | 0.618 | 0.222 | −1.042 | 1.775 |
|  | SD | 0.020 | 0.446 | 0.148 | 0.196 | 0.403 |

Lamb has the highest variability of expenditure elasticity with a Standard deviation (SD) value of 0.865. Lamb also has the highest province level expenditure elasticities mean value (1.606) that is followed by those of poultry (1.389), Pork (0.865) and beef (0.745). Except Lamb meat demand for which elasticities (Price and Expenditure Elasticities) values are in the interval of −6 and 5, the other meats elasticity values oscillate between −2 and 2 (Figure 3). Beef price variation induces more variation in meat demand/Consumption across the different provinces at the opposite of lamb price variation that induces in the study the lowest variation in meat demand (Figure 3).

## 4. Discussion

### The specification and the estimation method and results.

Income is one of the most important drivers of consumer preference at the retail level and one of the main explanatory variables for demand system estimation (Tryfos & Tryphonopoulos, 1973). Income is related to people (individuals) or households and is often defined at individual or/and household's level. Therefore, at the difference of territory level estimations, it could be interesting to estimate income at the household level in order to provide a breakdown of what is going across provinces taking account household characteristics. On the other hand, the elasticity values are in general affected by the estimation method and the demand specification (Gallet, 2010). To avoid problems related to the estimation method and the demand specification it is recommended by (Gallet, 2010) to use ready-made demand estimation systems for analysis as we did in that research. We used in this analysis a formal published ready-made demand estimation system thought R studio package "micEconAids" proposed by (Henningsen, 2017), which permits to perform quick and numerous analyses at Canada and its province level and have accurate results. Although several ready-made systems are sources of bias, it is not the case of the system used in this research that has been continually updated and has been scientifically proved to be accurate (Henningsen, 2017). Gallet (Gallet, 2010) state, that the location of demand and data characteristics have an impact on the demand estimation results. Our results highlight the scope of the impact of the location as well as the results of Komarek, Dunston et al. (Komarek et al., 2021) who examined in several locations a range of changes in the income elasticity of demand for red meat. Our research has revealed that the demand estimation result can be different with different interpretations in function of the location characteristics. In addition, it has been demonstrated that data quality plays a fundamental role in the plausibility of the demand estimation systems (Henningsen, 2017). In this research, an accurate dataset coming from a specialist of data collection led to a plausible demand estimation system and comforts the accuracy of our results. Hence, the methods and models used are plausible and in accordance with economic theories. Furthermore (Gallet, 2010) found that income elasticity is sensitive to a few factors, including data aggregation, and regional characteristics. Therefore, the location-based estimation of meat demand provides interesting information and good understanding of meat consumption patterns.

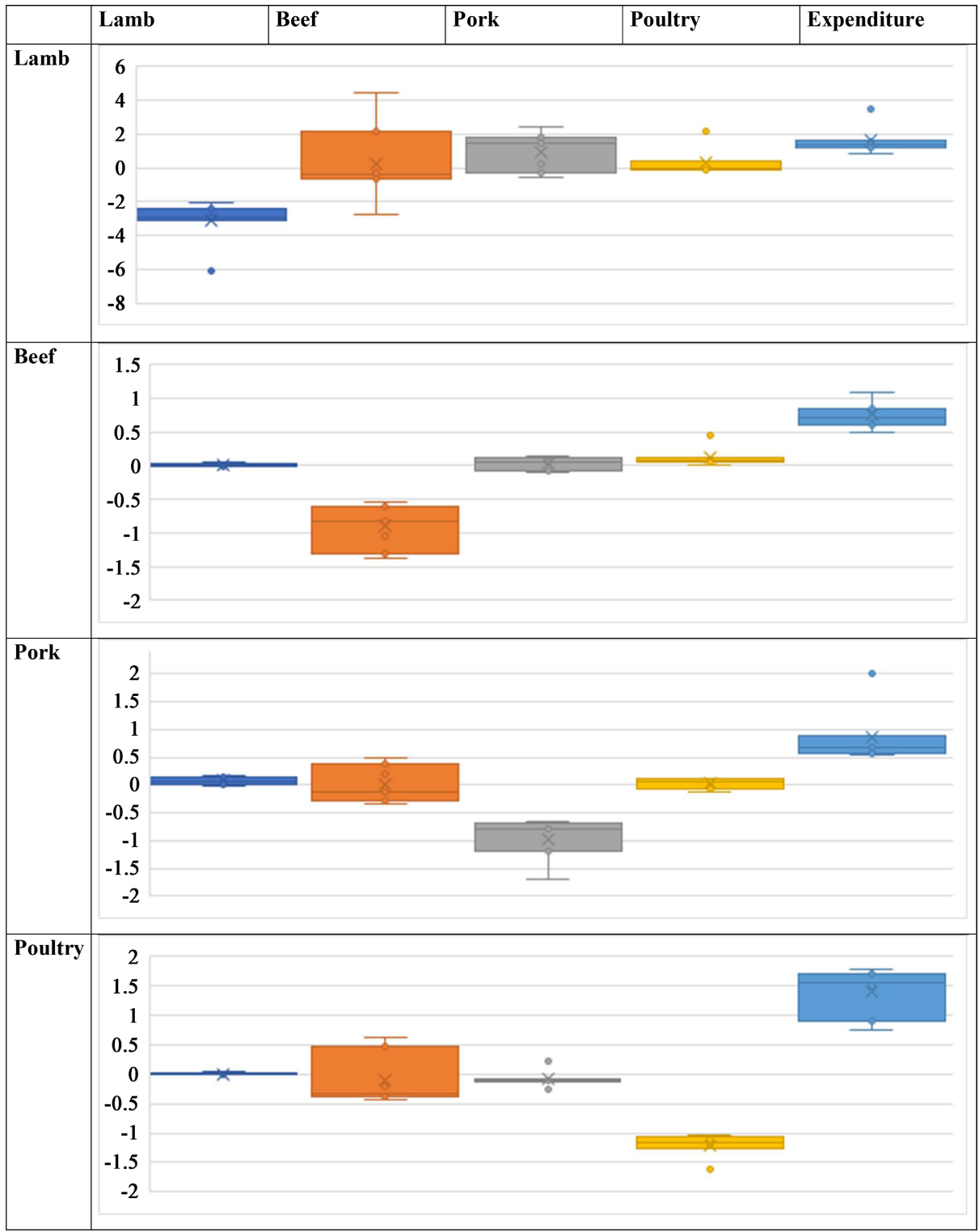

**Figure 3.** Marshallian (uncompensated) price and expenditure elasticities.

### Demand system estimation outcome

Demand for livestock-derived foods is undeniably dependent on factors such as income, price and consumer preferences, which interact with each other (Komarek et al., 2021). Meat consumption is also influenced by several factors, such as socio-economic, demographic, health, cultural and logistic constraints, availability, and climate change purpose. The model used and the different specifications in an estimation analysis can influence the elasticity values, and therefore the results may not be comparable. Anyway, our results are globally online with some authors results such as (Gallet, 2010) who found that meat demand price elasticities are in general lowest in the poultry case at the opposite of lamb meat price elasticity. This is the case of the own price elasticities in that research where lamb has the highest value and poultry the lowest values. Although they are two complementary products in Canada including all its provinces, poultry is less elastic compared to lamb. This is because poultry is cheaper, with low fat and cholesterol content, and suffers of no religious restrictions, thus considered as widely available healthy meat (Liu et al., 2022). The status of luxurious meat is confirmed for lamb in Canada and all its provinces, and that results are consistent across several studies. In addition to identifying the status of products using the expenditure/income elasticity as we did in our research, producers can use the income elasticity to assess the demand as income rises (Gallet, 2010) and design strategies. As we have done, several studies have estimated and analysed income elasticities as an important parameter. For example, Gallet (Gallet, 2010) conducted a meta-analysis focusing on income elasticities collected from 419 studies, including the following popular research (Blanciforti et al., 1986; Moschini & Meilke 1989; Asche & Wessells 1997).

To discuss the noteworthy results of the demand estimation, the complementarity between lamb and poultry meat and the status of luxury product of for lamb have been confirmed in European Union (EU) and the United Arab Emirates where lamb is qualified as natural meat, with a characteristic flavour and high price (Beriain et al., 2000; Basarir, 2013). The status of luxurious product for lamb is valid in all Provinces of Canada except in Alberta. Taking into account the location to estimate meat demand led to the simultaneous consideration of all subsequent factors that influence meet demand. The status of meat is thus different in function of the location and the magnitude of the meat con-sumption elasticity as well. Therefore, Lamb in addition of being a luxurious product is more sensible to the income in the province of Quebec compared to other provinces of Canada due to its high price like in EU (Beriain et al., 2000). A study on a meat consumption patterns in Australia have found that Beef is a luxury, while mutton, lamb, chicken, and pork are necessities (Wong et al., 2015). The difference of results in function of the location highlights the importance of exogen factors, including socio-economic and demographic characteristics not often included in a demand estimation. As in Australia, our study results classified Beef meat as luxurious product in the province of Québec while Beef is a necessity in all other provinces in Canada. Some particularities in QC an MS provinces reflected by the results are difficult to interpret, they need to be meticulously investigated for a good understanding. For example, Pork meat is classified as a luxurious product in the province of MS while pork is a necessity in all other provinces. In addition, Poultry is classified as a necessity both in QC and MS while it is a luxurious meat in the other provinces. Nonetheless, Except for poultry, other meats expenditure elasticity is higher in QC and MS which is consistent with poultry status of necessity in these locations. As 60.4% local poultry production produced in Quebec and Ontario, poultry status of necessity can be legitimate in Quebec.

### Socio-economic and demographic characteristic of the provinces

The world's population will reach 10 billion people by 2050, while aging and declining in some regions (Chan et al., 2021). This dynamic is clearly typic to Canada that integrates metropolitan areas with high demographic density and several territories that are sources of natural and agricultural resources but with the low demographic density. The difference in the demographic structure of Canada's territories plays an important role in the demand for meat, as some territories are dependent on economic activities and the availability of human resources. Demographic structure and economic development drive urbanisation and migration, which are linked to food demand and particularly meet demand. Due to the heterogeneity of Canada population repartition and the difference in economic activity distribution across the country, provinces such as Quebec (QC), Alberta (Al), British Columbia (BC), and Manitoba-Saskatchewan, and Manitoba (MS) have their specific tendencies of meat consumption. Consumers in these locations have uncommon behaviours due to their income levels, economic activities, and cultural and religious orientation. As Chan, Prager et al. (Chan et al., 2021) states in their study, consumer behaviour and the resulting meat demand affect food systems at the global or regional level. That is the case in locations with specific results, particularly Quebec and Alberta, which are two of the three main livestock producers in Canada. With the exception of Ontario, which is also one of the main livestock producing areas in Canada, Quebec and Alberta have a specific meat consumption pattern compared to the rest of Canada territory due to their lower dependence on logistics or trade. This leads to different meat consumption in the province thus a relative disparity of meat consumption across Canada and its provinces. By demonstrating that socio-economic and demographic characteristics, as well as household characteristics, play a critical role in understanding meat consumption patterns, this study provides a solid knowledge base for any research related to meat demand.

## 5. Conclusion

Analysis of Canada's national and disaggregated provincial meat consumption patterns has confirmed that locations impact meat demand system estimations. The research has revealed specific characteristics of provinces such as Quebec (Qc), Alberta (Al), and Manitoba-Saskatchewan, as well as Manitoba (MS) and British Columbia (BC) where results were sometimes different due to the specificities of these locations. The peculiarities of Quebec and Alberta are partly due to the fact that they are not subject to logistic and trade problems in terms of livestock base foods since they are part of the main livestock production area in Canada. Sometimes the differences were simply in the magnitude of the elasticity, highlighting the high sensibility of one location compared to another regarding the demand or status of a specific meat. The common conclusion for Canada and its provinces is the status of lamb as a luxury product, while the other three meats are necessity products across Canada. Meat consumption in each province is different, especially for Quebec (Qc), Manitoba-Saskatchewan and Manitoba (MS) and Alberta (Al) creating a disparity in meat demand patterns across Canada and its provinces. There is a need to meticulously investigate the sources of the disparity in meat demand patterns across Canada and its provinces, but also of improving the understanding of the causal relationships underlying lamb, beef, pork, and poultry consumption by considering individuals/households parallelly to locations.

## Acknowledgements

This research was supported by Laval University, Quebec sheep producers, Agri-Traçabilité Québec, the Centre D'expertise en Production Ovine du Québec (CEPOQ), the Centre D'études Des Coûts De Production en Agriculture (CECPA) and MITACS.

## Conflicts of Interest

The authors declare no conflicts of interest regarding the publication of this paper.

# Appendixes

## Appendix 1. Marshallian (Uncompensated) Price and Expenditure Elasticities (Canada)

| Quantities (Canada) | Marshallian (uncompensated) Price Elasticities | | | | Expenditure Elasticities |
|---|---|---|---|---|---|
| | lamb | beef | pork | poultry | |
| lamb | −2.639*** | 0.138 | 1.812 | −0.330** | 1.019 |
| beef | 0.005 | −0.650*** | −0.120** | −0.043. | 0.808*** |
| pork | 0.123 | −0.214* | −0.472** | 0.008 | 0.555*** |
| poultry | −0.016 | −0.392*** | −0.158*** | −0.932*** | 1.499*** |

"***" Significant at less than the 0.1% confidence level, "**" Significant at the 0.1% confidence level, "*" Significant at the 1% confidence level, "." Significant at the 5% confidence level, (blank) Significant at the 10% confidence level.

## Appendix 2. Marshallian (Uncompensated) Price and Expenditure Elasticities (Québec)

| Quantities (Québec) | Marshallian (uncompensated) Price Elasticities | | | | Expenditure Elasticities |
|---|---|---|---|---|---|
| | lamb | beef | pork | poultry | |
| lamb | −6.063** | 4.448 | −0.330 | 0.379 | 1.567 |
| beef | 0.044 | −1.383*** | 0.145** | 0.113. | 1.081*** |
| pork | −0.002 | 0.380*** | −1.198*** | −0.071 | 0.891 *** |
| poultry | 0.015 | 0.474* | −0.127 | −1.264*** | 0.902*** |

"***" Significant at less than the 0.1% confidence level, "**" Significant at the 0.1% confidence level, "*" Significant at the 1% confidence level, "." Significant at the 5% confidence level, (blank) Significant at the 10% confidence level.

## Appendix 3. Marshallian (Uncompensated) Price and Expenditure Elasticities (Maritimes)

| Quantities (Maritimes) | Marshallian (uncompensated) Price Elasticities | | | | Expenditure Elasticities |
|---|---|---|---|---|---|
| | lamb | beef | pork | poultry | |
| lamb | −2.890 | −0.390 | 1.793 | 0.107 | 1.380 |
| beef | −0.0008 | −0.827 *** | 0.043 | 0.066 | 0.719*** |
| pork | 0.072 | 0.196. | −0.722*** | −0.115. | 0.570*** |
| poultry | 0.0002 | −0.382*** | −0.267*** | −1.042*** | 1.691*** |

"***" Significant at less than the 0.1% confidence level, "**" Significant at the 0.1% confidence level, "*" Significant at the 1% confidence level, "." Significant at the 5% confidence level, (blank) Significant at the 10% confidence level.

## Appendix 4. Marshallian (Uncompensated) Price and Expenditure Elasticities (Ontario)

| Quantities (Ontario) | Marshallian (uncompensated) Price Elasticities | | | | Expenditure Elasticities |
|---|---|---|---|---|---|
| | lamb | beef | pork | poultry | |
| Lamb | −2.581 *** | −0.329 | 1.440 | −0.132 | 1.603 |
| Beef | 0.004 | −0.612*** | −0.047 | 0.054. | 0.601*** |
| Pork | 0.168* | −0.113 | −0.779*** | 0.120* | 0.605*** |
| poultry | −0.006 | −0.345*** | −0.116. | −1.105*** | 1.571*** |

"***" Significant at less than the 0.1% confidence level, "**" Significant at the 0.1% confidence level, "*" Significant at the 1% confidence level, "." Significant at the 5% confidence level, (blank) Significant at the 10% confidence level.

## Appendix 5. Marshallian (Uncompensated) Price and Expenditure Elasticities (Total West)

| Quantities (Total west) | Marshallian (uncompensated) Price Elasticities | | | | Expenditure Elasticities |
|---|---|---|---|---|---|
| | lamb | beef | pork | poultry | |
| Lamb | −2.900 ** | −0.693 | 2.434 | −0.011 | 1.170 |
| Beef | −0.010 | −0.604*** | −0.108*** | 0.006 | 0.716*** |
| Pork | 0.143. | −0.277** | −0.666*** | 0.119* | 0.681** |
| poultry | −0.004 | −0.385 *** | −0.090 | −1.068 *** | 1.547*** |

"***" Significant at less than the 0.1% confidence level, "**" Significant at the 0.1% confidence level, "*" Significant at the 1% confidence level, "." Significant at the 5% confidence level, (blank) Significant at the 10% confidence level.

## Appendix 6. Marshallian (Uncompensated) Price and Expenditure Elasticities (Alberta)

| Quantities (Alberta) | Marshallian (uncompensated) Price Elasticities | | | | Demand Elasticities |
|---|---|---|---|---|---|
| | lamb | beef | pork | poultry | |
| lamb | −3.099 | 2.174 | 0.251 | −0.144 | 0.819 |
| beef | 0.030 | −1.043*** | 0.098 * | 0.077** | 0.838*** |
| pork | 0.014 | 0.475*** | −1.108*** | 0.074. | 0.545*** |
| poultry | −0.008 | −0.219* | −0.114*** | −1.163*** | 1.504*** |

"***" Significant at less than the 0.1% confidence level, "**" Significant at the 0.1% confidence level, "*" Significant at the 1% confidence level, "." Significant at the 5% confidence level, (blank) Significant at the 10% confidence level.

## Appendix 7. Marshallian (Uncompensated) Price and Expenditure Elasticities (Man Sask)

| Quantities (Man Sask) | Marshallian (uncompensated) Price Elasticities | | | | Demand Elasticities |
|:---:|:---:|:---:|:---:|:---:|:---:|
| | lamb | beef | pork | poultry | |
| lamb | −2.407 | −2.714 | −0.526 | 2.174 | 3.473 |
| beef | −0.015 | −1.315*** | 0.123*** | 0.442*** | 0.765*** |
| pork | −0.006 | −0.273* | −1.697*** | −0.029 | 2.006*** |
| poultry | 0.043 | 0.618*** | 0.222*** | −1.618*** | 0.735** |

"***" Significant at less than the 0.1% confidence level, "**" Significant at the 0.1% confidence level, "*" Significant at the 1% confidence level, "." Significant at the 5% confidence level, (blank) Significant at the 10% confidence level.

## Appendix 8. Marshallian (Uncompensated) Price and Expenditure Elasticities (British Colombia)

| Quantities (British C) | Marshallian (uncompensated) Price Elasticities | | | | Demand Elasticities |
|:---:|:---:|:---:|:---:|:---:|:---:|
| | lamb | beef | pork | poultry | |
| lamb | −2.037* | −0.649 | 1.607 | −0.153 | 1.233 |
| beef | −0.009 | −0.537*** | −0.075. | 0.122*** | 0.494*** |
| pork | 0.146 | −0.345** | −0.685* | 0.127 | 0.756. |
| poultry | −0.014 | −0.441*** | −0.106 | −1.215*** | 1.775*** |

"***" Significant at less than the 0.1% confidence level, "**" Significant at the 0.1% confidence level, "*" Significant at the 1% confidence level, "." Significant at the 5% confidence level, (blank) Significant at the 10% confidence level.